\newcommand{\PRE}[1]{}       % Use if journal style
\documentclass[aps,prl,twocolumn,preprintnumbers,superscriptaddress,
amsmath,amssymb,showpacs,nofootinbib]{revtex4-1}
\newcommand{\ssection}[1]{{\em #1.\ }}
\usepackage{enumerate}
\usepackage{color}
\usepackage[pdftex]{graphicx}
\usepackage[colorlinks=true,citecolor=blue,urlcolor=blue]{hyperref}
\graphicspath{{paperfigures/}}	        % set directory for figures

\newcommand{\mev}{\text{MeV}}
\newcommand{\gev}{\text{GeV}}

\newcommand{\cm}{\text{cm}}
\newcommand{\m}{\text{m}}

\newcommand{\etal}{{\em et al.}}

\newcommand{\Eqsref}[2]{Equations~(\ref{#1}) and (\ref{#2})}
\renewcommand{\eqref}[1]{Eq.~(\ref{#1})}
\newcommand{\eqsref}[2]{Eqs.~(\ref{#1}) and (\ref{#2})}

\newcommand{\figref}[1]{Fig.~\ref{fig:#1}}

\newcommand{\tableref}[1]{Table~\ref{table:#1}}

\newcommand{\be}{{}^8\text{Be}}
\newcommand{\bezero}{\be}
\newcommand{\bestar}{\be^*}
\newcommand{\bestarprime}{\be^*{}'}

\begin{document}

\preprint{UCI-TR-2016-09}

\title{\PRE{\vspace*{1.5in}}
Protophobic Fifth Force Interpretation of the Observed Anomaly in $^8$Be Nuclear Transitions
\PRE{\vspace*{.5in}}}

\author{Jonathan L.~Feng\footnote{jlf@uci.edu}}
\affiliation{Department of Physics and Astronomy, University of
  California, Irvine, California 92697-4575 USA
\PRE{\vspace*{.2in}}}

\author{Bartosz Fornal\footnote{bfornal@uci.edu}}
\affiliation{Department of Physics and Astronomy, University of
  California, Irvine, California 92697-4575 USA
\PRE{\vspace*{.2in}}}

\author{Iftah Galon\footnote{iftachg@uci.edu}}
\affiliation{Department of Physics and Astronomy, University of
  California, Irvine, California 92697-4575 USA
\PRE{\vspace*{.2in}}}

\author{Susan Gardner\footnote{gardner@pa.uky.edu}}
\affiliation{Department of Physics and Astronomy, University of
  California, Irvine, California 92697-4575 USA
\PRE{\vspace*{.2in}}}
\affiliation{Department of Physics and Astronomy, University of 
Kentucky, Lexington, Kentucky 40506-0055 USA
\PRE{\vspace*{.4in}}}

\author{Jordan Smolinsky\footnote{jsmolins@uci.edu}}
\affiliation{Department of Physics and Astronomy, University of
  California, Irvine, California 92697-4575 USA
\PRE{\vspace*{.2in}}}

\author{Tim M.~P.~Tait\footnote{jttait@uci.edu}}
\affiliation{Department of Physics and Astronomy, University of
  California, Irvine, California 92697-4575 USA
\PRE{\vspace*{.2in}}}

\author{Philip Tanedo\footnote{flip.tanedo@uci.edu}}
\affiliation{Department of Physics and Astronomy, University of
  California, Irvine, California 92697-4575 USA
\PRE{\vspace*{.2in}}}

%\date{\today}

\begin{abstract}
\PRE{\vspace*{.2in}} Recently a 6.8$\sigma$ anomaly has been reported in the opening angle and invariant mass distributions of $e^+ e^-$ pairs produced in $\be$ nuclear transitions. The data are explained by a 17 MeV vector gauge boson $X$ that is produced in the decay of an excited state to the ground state, $\bestar \to \bezero\, X$, and then decays through $X \to e^+ e^-$.   The $X$ boson mediates a fifth force with a characteristic range of 12 fm and has milli-charged couplings to up and down quarks and electrons, and a proton coupling that is suppressed relative to neutrons.  The protophobic $X$ boson may also alleviate the current 3.6$\sigma$ discrepancy between the predicted and measured values of the muon's anomalous magnetic moment.
\end{abstract}

%14.70.Pw Other gauge bosons
% 27.20.+n	Properties of specific nuclei listed by mass ranges: 6 < A < 19
%21.30.-x	Nuclear forces
%12.60.Cn	Extensions of electroweak gauge sector
%13.60.-r	Photon and charged-lepton interactions with hadrons (for neutrino interactions, see 13.15.+g)

\maketitle

\ssection{Introduction}
\label{sec:intro}
The four known forces of nature, the electromagnetic, weak, strong, and gravitational interactions, are mediated by the photon, the $W$ and $Z$ bosons, the gluon, and the graviton, respectively.  The possibility of a fifth force, similarly mediated by an as-yet-unknown gauge boson, has been discussed~\cite{Lee:1955vk} since shortly after the introduction of Yang-Mills gauge theories, and has a rich, if checkered, history~\cite{Franklin:1993ki}.  If such a force exists, it must either be weak, or short-ranged, or both to be consistent with the wealth of experimental data.  In recent years, interest in this possibility has been heightened by the obvious need for dark matter, which has motivated new particles and forces in a dark or hidden sector that may mix with the visible sector and naturally induce a weak fifth force between the known particles.

Recently, studies of decays of an excited state of $\be$ to its ground state have found a 6.8$\sigma$ anomaly in the opening angle and invariant mass distribution of $e^+ e^-$ pairs produced in these transitions~\cite{Krasznahorkay:2015iga}.  The discrepancy from expectations may be explained by as-yet-unidentified nuclear reactions or experimental effects, but the observed distribution is beautifully fit by assuming the production of a new boson.  In this work, we advance the new particle interpretation, carefully considering the putative signal and the many competing constraints on its properties, and present a viable proposal for the new boson and the fifth force it induces.

\ssection{The $^8$Be Decay Anomaly}
\label{sec:Be}
The $\be$ nuclear excitation spectrum is precisely known~\cite{Tilley:2004zz}.  For this discussion, the most relevant $\be$ nuclear states and their properties are given in \tableref{states}.  To simplify our notation, we use the given symbols to denote specific states.  The ground state atomic mass is $8.005305\, \text{u}\simeq 7456.89\, \mev$; the ground state nuclear mass listed in \tableref{states} is about $4m_e$ below this.   There are also several unlisted broad resonance excited states both above and below $\bestar$ and $\bestarprime$ with widths as large as several MeV.

\begin{table}[tb]
 \caption{Relevant $\be$ states and their masses, decay widths, and spin-parity and isospin quantum numbers. }
 \vspace*{-.1in}
 \label{table:states}
\begin{tabular}[t]{lcccc}
 \toprule 
State &  \ Mass (MeV) \ & \ Width (keV) \ & \ $J^P$ \ & \ Isospin \\
\hline
 $\bestar$ (18.15) & 7473.00 & 138 & $1^+$ & 0 \\
  $\bestarprime$ (17.64) & 7472.49 & 10.7 & $1^+$ & 1 \\
   $\bezero$ (g.s.) & 7454.85 & --- & $0^+$ & 0 \\
\toprule
 \end{tabular}
\vspace*{-.2in}
\end{table}

In the experiment of Krasznahorkay \etal~\cite{Krasznahorkay:2015iga}, an intense proton beam impinges on thin $^7$Li targets.  Given the $^7$Li nucleus mass of 6533.83 MeV, the $\bestar$ and $\bestarprime$ states are resonantly produced by tuning the proton kinetic energies to 1.025 and 0.441 MeV, respectively.  The resulting excited states then decay promptly, dominantly back to $p\, ^7\text{Li}$, but also through rare electromagnetic processes.  For $\bestar$, radiative decay to the ground state has branching ratio $B(\bestar \to \bezero\, \gamma) \approx 1.4 \times 10^{-5}$, and there are also decays via internal pair conversion (IPC) with branching ratio $B(\bestar \to \bezero\, e^+ e^-) \approx 3.9 \times 10^{-3} B(\bestar \to \bezero\, \gamma) \approx 5.5 \times 10^{-8}$~\cite{Rose:1949zz}.  

For the IPC decays, one can measure the opening angle $\Theta$ between the $e^+$ and $e^-$ and also the invariant mass $m_{e^+ e^-}$.  One expects these distributions to be sharply peaked at low values of $\Theta$ and $m_{e^+ e^-}$ and fall smoothly and monotonically for increasing values.  This is not what is seen in the $\bestar$ decays.  Instead, there are pronounced bumps at $\Theta \approx 140^\circ$ and at $m_{e^+ e^-} \approx 17~\mev$~\cite{Krasznahorkay:2015iga}.  The experimental analysis fits the contributions from nearby broad resonances, but these cannot reproduce the shape of the observed excesses.  The deviation has a significance of 6.8$\sigma$, corresponding to a background fluctuation probability of $5.6\times 10^{-12}$~\cite{Krasznahorkay:2015iga}. The excess is maximal on the $\bestar$ resonance and disappears as the proton beam energy is moved off resonance.  No such effect is seen in $\bestarprime$ IPC decays.

The fit may be improved by postulating a new boson $X$ that is produced on-shell in $\bestar \to \bezero\, X$ and decays promptly via $X \to e^+ e^-$.  The authors of Ref.~\cite{Krasznahorkay:2015iga} have simulated this process, including the detector energy resolution, which broadens the $m_{e^+ e^-}$ peak significantly~\cite{Gulyas:2015mia}.  They find that the observed excess's shape and size are beautifully fit by a new boson with mass $m_X = 16.7\pm 0.35\, \text{(stat)} \pm 0.5\, \text{(sys)}~\mev$ and relative branching ratio $B(\bestar \to \bezero\, X) / B(\bestar \to \bezero\, \gamma) = 5.8 \times 10^{-6}$, assuming $B(X \to e^+ e^-) = 1$. With these values, the fit had a $\chi^2/\text{dof} = 1.07$.

\ssection{Protophobic Gauge Bosons}
\label{sec:5thForce}
{\em A priori} the $X$ boson may be a scalar, pseudoscalar, vector, axial vector, or even a spin-2 particle.  Some of these cases are easy to dismiss.  If parity is conserved, the $X$ boson cannot be a scalar: in a $1^+ \to 0^+ 0^+$ transition, angular momentum conservation requires the final state to have $L = 1$, but parity conservation requires $+1 = (-1)^L$.  Decays to a pseudoscalar $0^-$ state are not forbidden by any symmetry, but are severely constrained by experiment.  For such axion-like particles $a$, the two-photon interaction $g_{a \gamma \gamma} a F^{\mu\nu} \tilde{F}_{\mu\nu}$ is almost certainly present at some level, but for $m_a \approx 17~\mev$, all coupling values in the range $ 1 / (10^{18}~\gev) < g_{a \gamma \gamma} < 1/ (10~\gev)$ are excluded~\cite{Hewett:2012ns,Dobrich:2015jyk}.

Here we focus on the vector case.  We consider a massive spin-1 Abelian gauge boson $X$ that couples non-chirally to standard model (SM) fermions with charges $\varepsilon_f$ in units of $e$.  The new Lagrangian terms are
\begin{equation}
{\cal L} = - \frac{1}{4} X_{\mu\nu} X^{\mu\nu} + \frac{1}{2} m_X^2 X_\mu X^\mu - X^{\mu} J_{\mu} , 
\end{equation}
 where $X$ has field strength $X_{\mu\nu}$ and couples to the current  $J_{\mu} = \sum_f e \varepsilon_f \bar{f} \gamma_{\mu} f$, or, at the nucleon level, $J^N_{\mu} = e \varepsilon_p \bar{p} \gamma_{\mu} p + e \varepsilon_n \bar{n} \gamma_{\mu} n$, with $\varepsilon_p = 2\varepsilon_u + \varepsilon_d$ and $\varepsilon_n = \varepsilon_u + 2 \varepsilon_d$.

We first determine what values of the charges are required to fit the $\be$ signal.
The characteristic energy scale of the decay $\bestar \to \bezero\, X$ is 10 MeV, and so we may consider an effective theory in which $\bestar$, $\bezero$, and $X$ are the fundamental degrees of freedom.  The one effective operator consistent with the $J^P$ quantum numbers of these states is 
\begin{equation}
{\cal L}_{\text{int}} = \frac{1}{\Lambda} \epsilon^{\mu\nu\alpha\beta} 
\left( \partial_{\mu} \bestar_{\nu} - \partial_{\nu} \bestar_{\mu} \right)
X_{\alpha\beta} \bezero \ .
\end{equation}
The matrix element $\langle \bezero X | {\cal L}_{\text{int}} | \bestar \rangle$ is proportional to $\langle \bezero | J^N_{\mu} | \bestar \rangle = (e /2) (\varepsilon_p + \varepsilon_n) {\cal M}$, where ${\cal M} = 
\langle \bezero | (\bar{p} \gamma_{\mu} p + \bar{n} \gamma_{\mu} n) | \bestar \rangle$ contains the isoscalar component of the current, since the initial and final states are both isoscalars.  The resulting decay width is
\begin{equation}
\Gamma (\bestar \to \bezero\, X) = \frac{(e/2)^2 (\varepsilon_p + \varepsilon_n)^2 }{3 \pi \Lambda^2} | {\cal M} | ^2 
| \vec{p}_X|^3 \ .
\end{equation}
To fit the signal, we need
\begin{equation}
\frac{B ( \bestar \to \bezero \, X )}{B ( \bestar \to \bezero \, \gamma)}
= (\varepsilon_p \! + \varepsilon_n)^2 \frac{|\vec{p}_X|^3}{|\vec{p}_{\gamma}|^3} \approx 5.8 \times 10^{-6} ,
\end{equation}
where, up to higher-order corrections~\cite{Pastore:2014oda}, both the nuclear matrix elements and the scale $\Lambda$ have canceled in the ratio.
For $m_X = 17~\mev$, we require $| \varepsilon_p + \varepsilon_n | \approx 0.011$, or
\begin{equation}
| \varepsilon_u + \varepsilon_d | \approx 3.7 \times 10^{-3} \ .
\label{quarkcharges}
\end{equation}

The 17 MeV $X$ boson is produced through hadronic couplings, but can decay only to $e^+ e^-$, $\nu \bar{\nu}$, or $\gamma \gamma \gamma$.  (We assume there are no decays to unknown particles.) The three-photon decay is negligible, and we will assume that decays to neutrinos are also highly suppressed, for the reasons given below.  The $X$ boson then decays through its electron coupling with width~\cite{Pospelov:2008zw}
\begin{equation}
\Gamma (X \to e^+ e^-) = \varepsilon_e^2 \alpha \frac{m_X^2 + 2 m_e^2}{3 m_X} 
\sqrt{1 - 4 m_e^2 / m_X^2 } \ .
\end{equation}
The $X$ boson is produced with velocity $v \approx 0.35c$ in the $\bestar$ frame, which is moving non-relativistically with $v = 0.017c$ relative to the lab frame.  The $X$ mean decay length is $L \approx \varepsilon_e^{-2} \, 1.8 \times 10^{-12}~\m$ in the lab frame.  The $X$ boson must decay promptly in the experimental setup of Refs.~\cite{Krasznahorkay:2015iga,Gulyas:2015mia} so that the $e^+ e^-$ decay products are detected and the $\Theta$ measurements are not distorted.  Requiring $L \alt 1~\cm$, for example, implies 
\begin{equation}
| \varepsilon_e |  \agt 1.3 \times 10^{-5} \ .
\label{electroncharge}
\end{equation} 

{}From \eqref{quarkcharges}, we see that a dark photon cannot explain the $\be$ anomaly.  For a dark photon, fermions have charges proportional to their SM charges, $\varepsilon_f = q_f \varepsilon$, where $\varepsilon$ is the kinetic mixing parameter, and so \eqref{quarkcharges} implies $\varepsilon \approx 0.011$. This is excluded by many experiments, and most stringently by the NA48/2 experiment, which requires $\varepsilon < \varepsilon_{\text{max}} = 8 \times 10^{-4}$ at 90\% CL~\cite{Batley:2015lha}. The authors of Ref.~\cite{Krasznahorkay:2015iga} estimated that $\varepsilon^2 \sim 10^{-7}$ can fit the signal, but this value of $\varepsilon$ is far too small, in part because of  the $| \vec{p}|^3$ suppression of the signal.

The NA48/2 bound, however, does not exclude a general vector boson interpretation of the $\be$ anomaly.  The NA48/2 limit is a bound on $\pi^0 \to X \gamma$.  In the general gauge boson case, this is proportional to the anomaly trace factor $N_\pi \equiv (\varepsilon_u q_u - \varepsilon_d q_d)^2$.  Applying the dark photon bound $N_{\pi} < \varepsilon_{\text{max}}^2 / 9$, we find that, for a general gauge boson, 
\begin{equation}
|2 \varepsilon_u + \varepsilon_d | < \varepsilon_{\text{max}} = 8 \times 10^{-4} \ .
\label{pion}
\end{equation}
\Eqsref{quarkcharges}{pion} may be satisfied with a mild $\sim 10\%$ cancelation, provided the charges satisfy
\begin{equation}
-2.3 < \frac{\varepsilon_d}{\varepsilon_u} < -1.8 \ ,
\quad -0.067 < \frac{\varepsilon_p}{\varepsilon_n} < 0.078 \ .
\end{equation}
Given the latter condition, we call the general class of vector models that can both explain the $\be$ anomaly and satisfy pion decay constraints ``protophobic."

\begin{figure}[t]
\includegraphics[width=.92\linewidth]{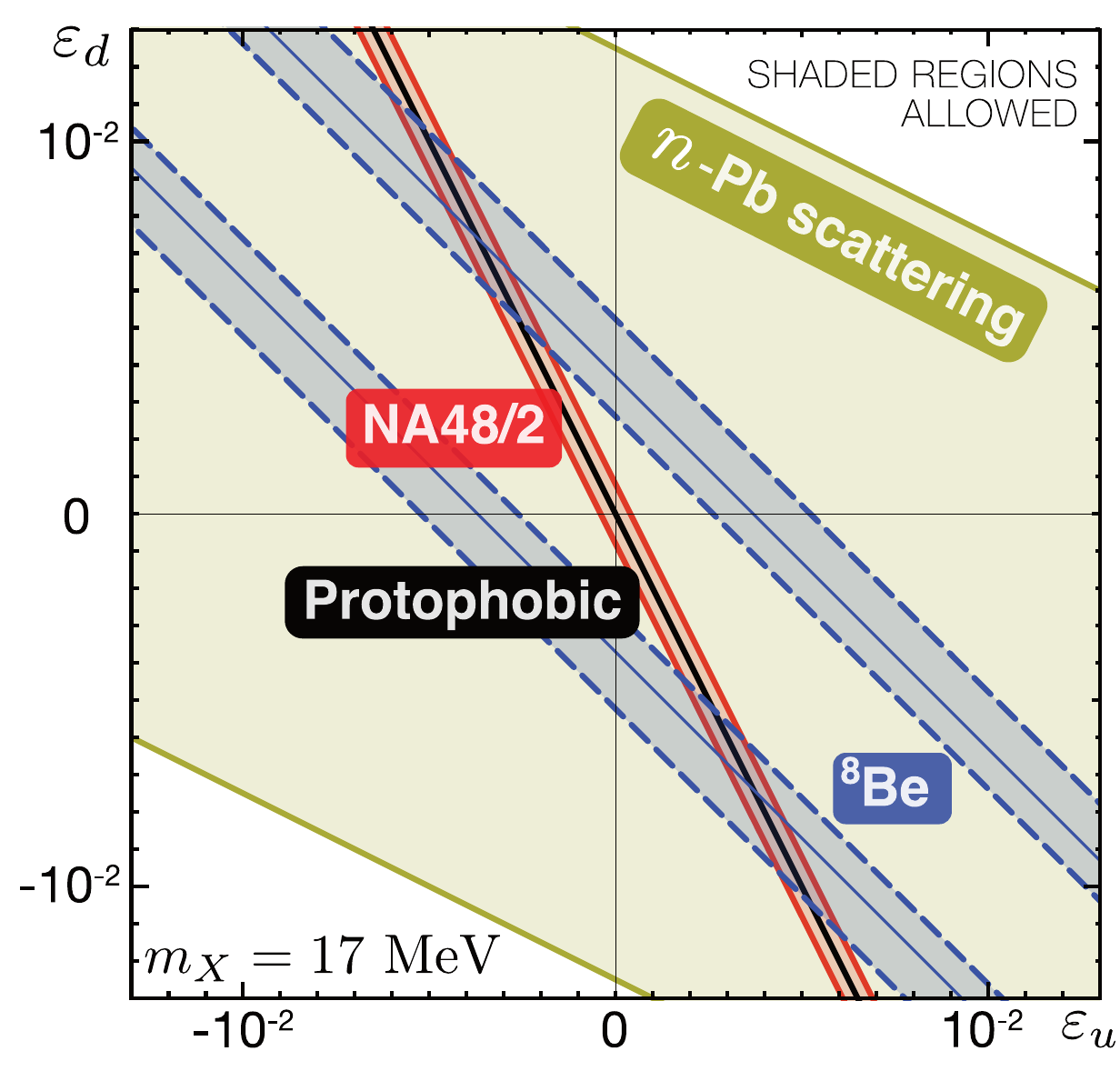} \\
\includegraphics[width=.92\linewidth]{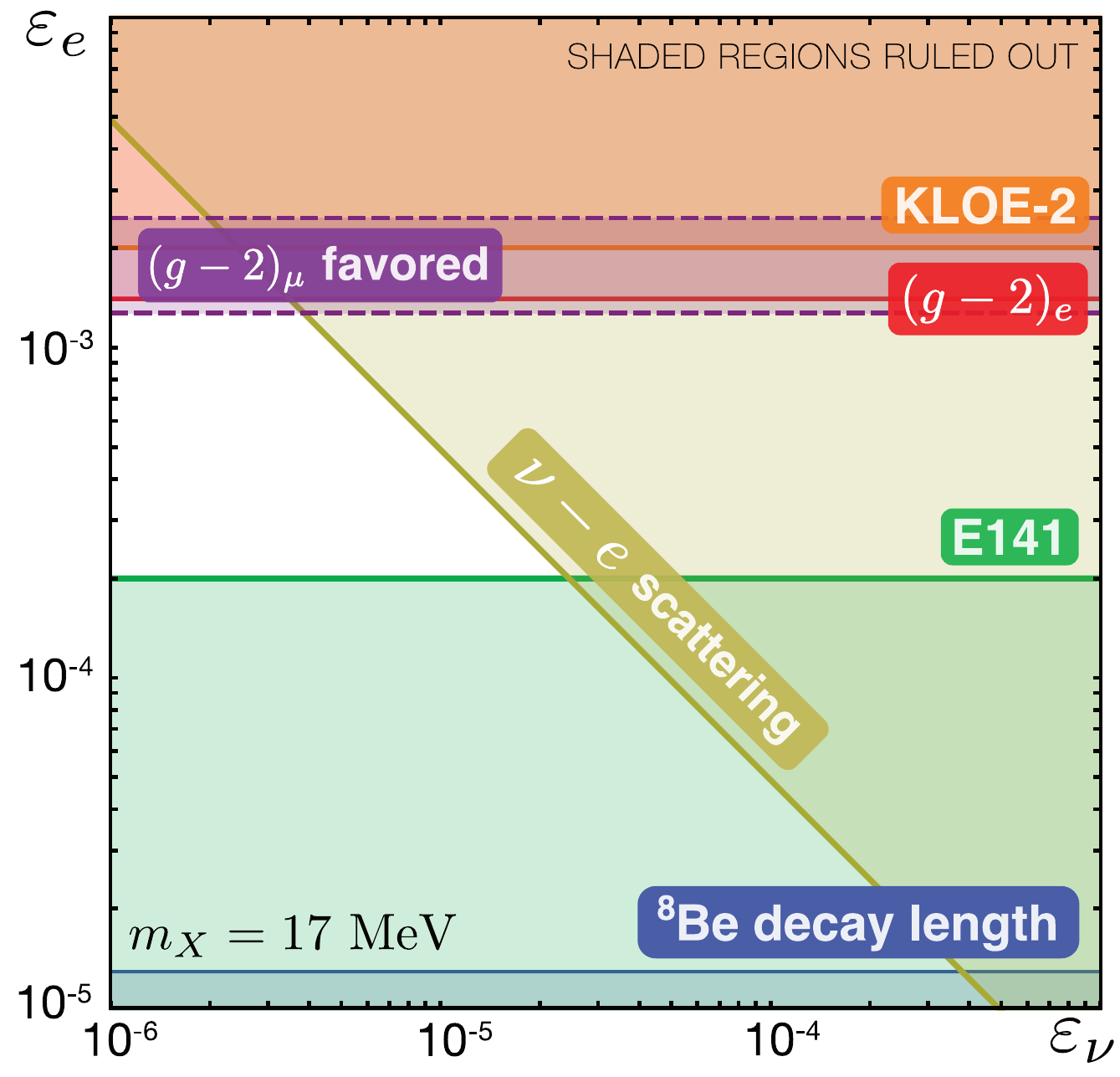}
\vspace*{-0.1in}
\caption{The required charges to explain the $\be$ anomaly in the $(\varepsilon_u, \varepsilon_d)$ (top) and $(\varepsilon_{\nu}, \varepsilon_e)$ (bottom) planes, along with the leading constraints discussed in the text. {\em Top}: The $n$-Pb and NA48/2 constraints are satisfied in the shaded regions. On the protophobic contour, $\varepsilon_d / \varepsilon_u = -2$.  The width of the $\be$ bands corresponds to requiring the signal strength to be within a factor of 2 of the best fit.  {\em Bottom}: The E141, KLOE-2, $(g-2)_e$, and $\nu-e$ scattering constraints exclude their shaded regions, whereas $(g-2)_{\mu}$ favors its shaded region.  The $\be$ signal imposes a lower bound on $| \varepsilon_e |$.  
\label{fig:limits} }
\vspace*{-0.2in}
\end{figure}

\ssection{Constraints from Other Experiments}
\label{sec:constraints}
Although there is no need for the gauge boson to decouple from protons completely, for simplicity, for the rest of this work, we consider the extreme protophobic limit where $\varepsilon_p = 0$.  We parametrize the quark charges as 
$\varepsilon_u = - \frac{1}{3} \varepsilon_n$, 
$\varepsilon_d = \frac{2}{3} \varepsilon_n$ and determine what choices for $\varepsilon_n$, $\varepsilon_e$, and $\varepsilon_{\nu}$ are viable. We focus on these first-generation charges, as the $\be$ signal depends on them, but include comments on the charges of the other generations below. The charges required to explain the $\be$ signal, along with the leading bounds discussed below, are shown in \figref{limits}.

As noted above, the decay $\bestarprime \to \bezero \, X$ is not seen.  The protophobic gauge boson can mediate isovector transitions, so there is no dynamical suppression of this decay. However, its mass is near the 17.64 MeV threshold, so the decay is  kinematically suppressed.  For $m_X = 17.0~(17.4)~\mev$, the $|\vec{p}_X|^3 / |\vec{p}_{\gamma}|^3$ phase space suppression factor is 2.3 (5.2) times more severe for the $\bestarprime$ decay than for the $\bestar$ decay.  In particular, $m_X = 17.4~\mev$ is within $1\sigma$ of the central value, and a 5.2 times smaller signal in the $\bestarprime$ decay is consistent with the data. We will continue to refer to the boson as a 17 MeV boson, as no other processes are sensitive to the precise value of its mass, with the understanding that the null $\bestarprime$ result may require it to be a bit above 17 MeV.  Note that although $m_X = 17.4~\mev$ is near the endpoint of the $\bestarprime$ decay, it is not near the endpoint of the $\bestar$ decay, and the $\Theta$ and $m_{e^+ e^-}$ distributions return to near their SM values at high values. This is not a ``last bin" effect.  

A number of experiments provide upper bounds on $| \varepsilon_e |$. The anomalous magnetic moment of the electron, $(g-2)_e$, constrains $|\varepsilon_e| < 1.4 \times 10^{-3}$ (3$\sigma$)~\cite{Davoudiasl:2014kua}. The KLOE-2 experiment has looked for $e^+ e^- \to \gamma X$, followed by $X \to e^+ e^-$, and finds $| \varepsilon_e | < 2 \times 10^{-3}$~\cite{Anastasi:2015qla}.  A similar search at BaBar has reached similar sensitivity in $\varepsilon_e$, but is limited to $m_X > 20~\mev$~\cite{Lees:2014xha}. 

Electron beam dump experiments also constrain $\varepsilon_e$ by searching for $X$ bosons radiated off electrons that scatter on target nuclei.  As a group, these exclude $| \varepsilon_e |$ in the $10^{-8}$ to $10^{-4}$ range~\cite{Essig:2013lka}.  For this discussion, given \eqref{electroncharge}, these experiments provide lower bounds on $|\varepsilon_e|$.  In more detail, for $m_X \approx 17~\mev$, SLAC experiment E141 requires $|\varepsilon_e| > 2 \times 10^{-4}$~\cite{Riordan:1987aw,Bjorken:2009mm}.  There are also less stringent bounds from Orsay~\cite{Davier:1989wz} and SLAC's E137~\cite{Bjorken:1988as} and Millicharge~\cite{Diamond:2013oda} experiments, and Fermilab experiment E774~\cite{Bross:1989mp} excludes some couplings when $m_X < 10~\mev$.  

We now turn to bounds on the hadronic couplings.  We have already discussed the NA48/2 bound from $\pi^0$ decays.  WASA-at-COSY has also published a bound based on $\pi^0$ decays, but it is weaker and applies only for $m_X > 20~\mev$~\cite{Adlarson:2013eza}.   Potentially more problematic is a bound from the HADES experiment, which searches for $X$ bosons in $\pi^0$, $\eta$, and $\Delta$ decays and excludes the dark photon parameter $\varepsilon \agt 3 \times 10^{-3}$, but this also applies only for $m_X > 20~\mev$~\cite{Agakishiev:2013fwl}.  Note also that $\pi^0 \to X X \to e^+ e^- e^+ e^-$ is not suppressed by the protophobic charge assignments, but it is suppressed by $\varepsilon_n^4$ and, for $| \varepsilon_n | \sim 10^{-2}$, this is below current sensitivities.  Similar considerations suppress $X$ contributions to other decays, such as $\pi^+ \to \mu^+ \nu_{\mu} e^+ e^-$, to acceptable levels.  

The hadronic charge can also be bounded by limits on Yukawa potentials from neutron-nucleus scattering.  For a Yukawa potential $-g_n^2 A e^{-m_X r} / (4 \pi r)$, $n$--Pb scattering requires $g_n^2 /(4 \pi) < 3.4 \times 10^{-11} (m_X / \mev)^4$~\cite{Barbieri:1975xy}.  The protophobic $X$ boson induces a Yukawa potential $\varepsilon_n^2 \alpha (A-Z) e^{-m_X r} / r$.  Given $Z = 82$ and $A=208$ for Pb, the bounds imply $| \varepsilon_n | < 2.5 \times 10^{-2}$. 

There are constraints from proton fixed target experiments.  The $\nu$-Cal I experiment at the U70 accelerator at IHEP provides a well-known dark photon constraint, but its bounds are derived from $X$-bremsstrahlung from the initial $p$ beam and $\pi^0$ decays to $X$ bosons~\cite{Blumlein:2013cua}.  Both of these are suppressed in protophobic models.  The CHARM experiment at CERN also bounds the parameter space through searches for $\eta, \eta' \to X \gamma$, followed by $X \to e^+e^-$~\cite{Gninenko:2012eq}.  At the upper boundary of the region excluded by CHARM, the constraint is determined almost completely by the parameters that enter the $X$ decay length, and so the dark photon bound on $\varepsilon$ applies to $\varepsilon_e$ and requires $| \varepsilon_e | > 2 \times 10^{-5}$.  A similar, but weaker constraint can be derived from LSND data~\cite{Athanassopoulos:1997er,Batell:2009di,Essig:2010gu}.

There are also bounds on the neutrino charge $\varepsilon_{\nu}$.  In the present case, where $\varepsilon_e$ is non-zero, a recent study of $B-L$ gauge bosons~\cite{Bilmis:2015lja} finds that these couplings are most stringently bounded by precision studies of $\bar{\nu} - e$ scattering from the Taiwan Experiment on Neutrinos (TEXONO) for the $m_X$ of interest here~\cite{Deniz:2009mu}.  Reinterpreted for the present case, these studies require $| \varepsilon_{\nu} \varepsilon_e|^{1/2} \alt  7 \times 10^{-5}$.   There are also bounds from coherent neutrino-nucleus scattering.  Dark matter experiments with Xe target nuclei require a $B-L$ gauge boson to have coupling $g_{B-L} \alt 4 \times 10^{-5}$~\cite{Cerdeno:2016sfi}.  Rescaling this to the current case, given $Z = 54$ and $A= 131$ for Xe, we find $| \varepsilon_{\nu} \varepsilon_n |^{1/2} < 2 \times 10^{-4}$.  

To explain the $\be$ signal, $\varepsilon_n$ must be significantly larger than $\varepsilon_e$.  Nevertheless, the $\bar{\nu} - e$ scattering constraint provides a bound on $\varepsilon_{\nu}$ that is comparable to or stronger than the $\nu - N$ constraint throughout parameter space, and so we use the $\bar{\nu} - e$ constraint below.  Note also that, given the range of acceptable $\varepsilon_e$, the bounds on $\varepsilon_{\nu}$ are more stringent than the bounds on $\varepsilon_e$, and so $B(X \to e^+ e^-) \approx 100\%$, justifying our assumption above.

Although not our main concern, there are also bounds on second-generation couplings.  For example, NA48/2 also derives bounds on $K^+ \to \pi^+ X$, followed by $X \to e^+ e^-$~\cite{Batley:2015lha}.  However, this branching ratio vanishes for massless $X$ and is highly suppressed for low $m_X$.  For $m_X = 17~\mev$, the bound on $\varepsilon_n$ is not competitive with those discussed above~\cite{Pospelov:2008zw,Davoudiasl:2014kua}.  The KLOE-2 experiment also searches for $\phi \to \eta X$ followed by $X \to e^+ e^-$ and excludes the dark photon parameter $\varepsilon \agt 7 \times 10^{-3}$~\cite{Babusci:2012cr}. This is similar numerically to bounds discussed above, and the strange quark charge $\varepsilon_s$ can be chosen to satisfy this constraint.  

In summary, in the extreme protophobic case with $m_X \approx 17~\mev$, the charges are required to satisfy $| \varepsilon_n | < 2.5 \times 10^{-2}$ and $2 \times 10^{-4} < | \varepsilon_e | < 1.4\times 10^{-3}$, and $| \varepsilon_{\nu} \varepsilon_e|^{1/2} \alt  7 \times 10^{-5}$.  
Combining these with \eqsref{quarkcharges}{electroncharge}, we find that
a protophobic gauge boson with first-generation charges 
\begin{align}
\varepsilon_u = - \frac{1}{3} \varepsilon_n \approx \pm 3.7 \times 10^{-3} \nonumber \\
\varepsilon_d = \frac{2}{3} \varepsilon_n \approx \mp 7.4 \times 10^{-3} \nonumber \\
2 \times 10^{-4} \alt | \varepsilon_e |  \alt 1.4 \times 10^{-3} \nonumber
\\
\left| \varepsilon_{\nu} \varepsilon_e \right|^{1/2} \alt 7 \times 10^{-5} 
\label{erange} 
\end{align} 
explains the $\be$ anomaly by $\bestar \to \bezero \, X$, followed by $X \to e^+ e^-$, consistent with existing constraints. For $| \varepsilon_e |$ near the upper end of the allowed range in \eqref{erange} and $| \varepsilon_{\mu} | \approx | \varepsilon_e |$, the $X$ boson also solves the $(g-2)_\mu$ puzzle, reducing the current 3.6$\sigma$ discrepancy to below 2$\sigma$~\cite{Pospelov:2008zw}.

\ssection{Conclusions}
\label{sec:conclusions}
We find evidence in the recent observation of a 6.8$\sigma$ anomaly in the $e^+ e^-$ distribution of nuclear $\be$ decays for a new vector gauge boson.  The new particle mediates a fifth force with a characteristic length scale of 12 fm.  The requirements of the signal, along with the many constraints from other experiments that probe these low energy scales, constrain the mass and couplings of the boson to small ranges: its mass is $m_X \approx 17~\mev$, and it has milli-charged couplings to up and down quarks and electrons, but with relatively suppressed (and possibly vanishing) couplings to protons (and neutrinos) relative to neutrons.  If its lepton couplings are approximately generation-independent, the 17 MeV vector boson may simultaneously explain the existing 3.6$\sigma$ deviation from SM predictions in the anomalous magnetic moment of the muon.  
It is also interesting to note that couplings of this magnitude, albeit in an axial vector case, may resolve a 3.2$\sigma$ excess in $\pi^0 \to e^+ e^-$ decays~\cite{Abouzaid:2006kk,Kahn:2007ru}.

To confirm the $\be$ signal, the most similar approach would be to look for other nuclear states that decay to discrete gamma rays with energies above 17 MeV through M1 or E1 electromagnetic transitions.  Unfortunately, the $\be$ system is quite special and the $\bestar$ and $\bestarprime$ states yield gamma rays that are among the most energetic of all the nuclear states~\cite{nndc}.

\begin{figure}[t]
\includegraphics[width=.92\linewidth]{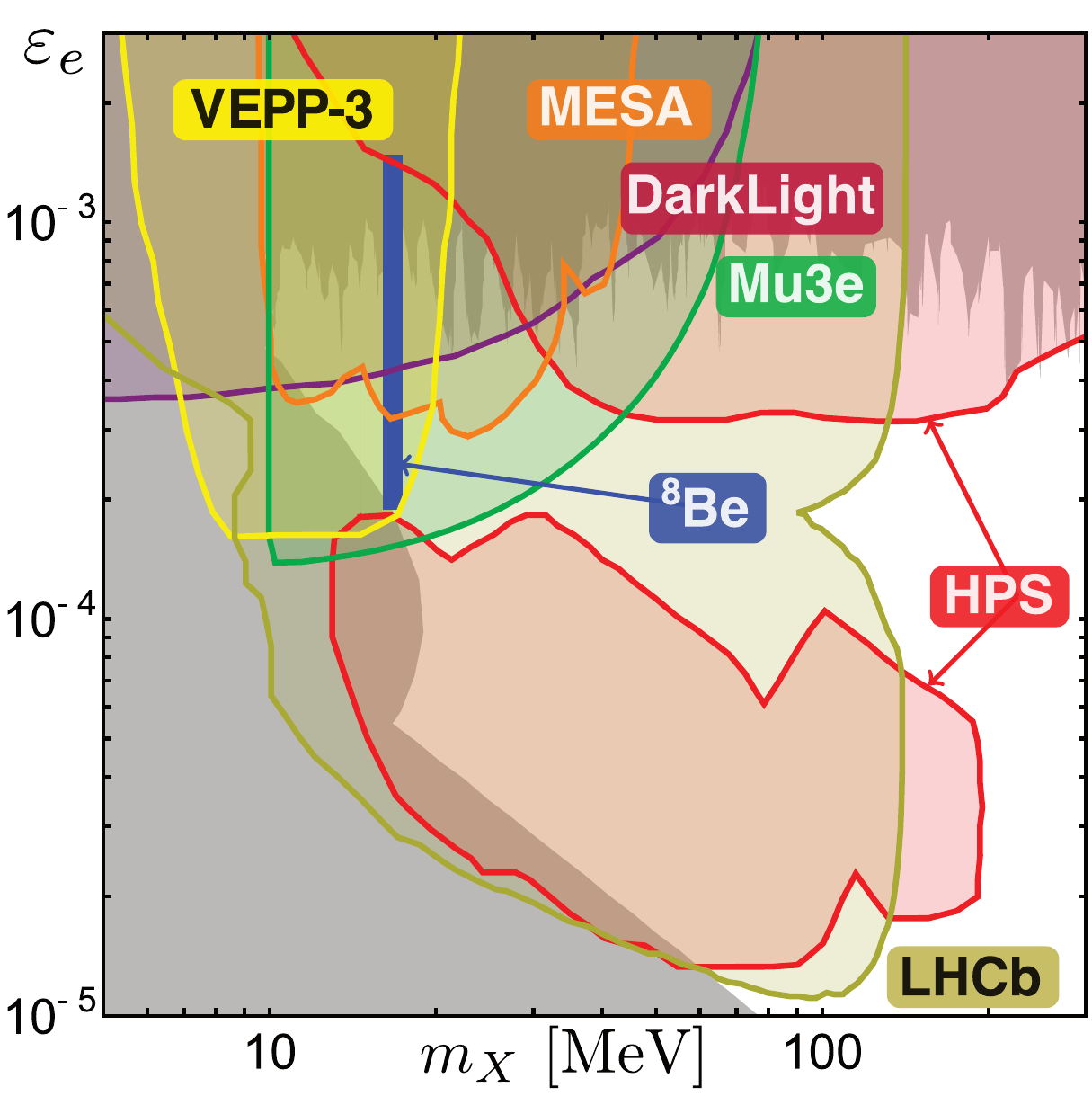} 
\vspace*{-0.1in}
\caption{The $\be$ signal region, along with current constraints discussed in the text (gray) and projected sensitivities of future experiments in the $(m_X, \varepsilon_e)$ plane.  For the $\be$ signal, the other couplings are assumed to be in the ranges given in \eqref{erange}; for all other contours, the other couplings are those of a dark photon. 
\label{fig:future} }
\vspace*{-0.2in}
\end{figure}

Nevertheless there are myriad opportunities to test and confirm this explanation, including re-analysis of old data sets, ongoing experiments, and many planned and future experiments, including DarkLight~\cite{Balewski:2014pxa}, HPS~\cite{Moreno:2013mja}, LHCb~\cite{Ilten:2015hya}, MESA~\cite{Beranek:2013yqa}, Mu3e~\cite{Echenard:2014lma}, VEPP-3~\cite{Wojtsekhowski:2012zq}, and possibly also SeaQuest~\cite{Gardner:2015wea} and SHiP~\cite{Anelli:2015pba}. The $\be$ signal region and expected sensitivities of these experiments are shown in \figref{future}. It will also be important to embed the protophobic gauge boson in UV-complete extensions of the standard model, a task made challenging by the wealth of data constraining new physics at the $\sim 10~\mev$ energy scale.  Further details about the existing constraints, prospects for the future, and UV completions will be presented elsewhere~\cite{Feng:2016ysn}.

\ssection{Acknowledgments}
We thank Attila J.~Krasznahorkay, Alexandra Gade, and Alan Robinson for helpful correspondence.  The work of J.L.F., B.F., I.G., J.S., T.M.P.T., and  P.T.\ is supported in part by  NSF Grant No.~PHY-1316792. The work of S.G. is supported in part by the DOE Office of Nuclear Physics under contract DE-FG02-96ER40989.  J.L.F. is supported in part by a Guggenheim Foundation grant and in part by Simons Investigator Award \#376204.

\bibliography{5thForce2}

\end{document}